\documentclass{ws-rv9x6}
\usepackage{subfigure}     

\usepackage[square]{ws-rv-van}     
\usepackage{ws-index}     
\makeindex
\newindex{aindx}{adx}{and}{Author Index}       
\renewindex{default}{idx}{ind}{Subject Index}  

\usepackage{dsfont,color}

 \def\e     {\mathrm{e}}

 \def\sign  {\operatorname{sign}}

 \renewcommand{\Pr    }[1]{\mathds{P}\!\left[#1\right]}
   \newcommand{\Exp   }[1]{      \exp\!\left(#1\right)}
   \newcommand{\Erfc  }[1]{\operatorname{erfc}\!\left(#1\right)}
   \newcommand{\esp   }[1]{\mathds{E}[#1]}
   \newcommand{\var   }[1]{\mathds{V}[#1]}
   \newcommand{\pr    }[1]{\mathds{P}[#1]}
   \newcommand{\1     }[1]{\mathds{1}_{\{#1\}}}
 \renewcommand{\d     }[1]{\mathrm{d}#1} 
 
 \newcommand{\FF}[3]{{\vphantom{#2}}#1#2#3}

  \newcommand{\vect}[1]{\boldsymbol{#1}} 

\def\argmax{\operatornamewithlimits{arg\,max}}
\def\argmin{\operatornamewithlimits{arg\,min}}

\newcommand{\qstar}{{q^*}}

\begin{document}

\chapter[Some applications of first-passage ideas to finance]{Some applications of first-passage ideas to finance}

\author[R. Chicheportiche, J.-P. Bouchaud]{R\'emy Chicheportiche, Jean-Philippe Bouchaud}
\index[aindx]{Bouchaud, J.-P.} 
\index[aindx]{Chicheportiche, R.}

\address{Capital Fund Management,\\
23--25, rue de l'Universit\'e, 75\,007 Paris}

\begin{abstract}
Many problems in finance are related to first passage times. 
Among all of them, we chose three on which we contributed personally. 
Our first example relates Kolmogorov-Smirnov like goodness-of-fit tests, 
modified in such a way that tail events and core events contribute equally to the test 
(in the standard Kolmogorov-Smirnov, the tails contribute very little to the measure of goodness-of-fit). 
We show that this problem can be mapped onto that of a random walk inside moving walls. 
The second example is the optimal time to sell an asset (modelled as a random walk with drift) 
such that the sell time is as close as possible to the time at which the asset reaches its maximum value. 
The last example concerns optimal trading in the presence of transaction costs. 
In this case, the optimal strategy is to wait until the predictor reaches (plus or minus) a threshold value 
before buying or selling. The value of this threshold is found by mapping the problem onto that 
of a random walk between two walls. 
\end{abstract}
\body

\clearpage
\section{Introduction}

Quantitative finance is the bounty land of statistics and probabilities. 
Bachelier proposed to model price paths as random walks in 1900, 
but his amazingly creative work was forgotten until the sixties, 
when Samuelson, Black \& Scholes revived the Brownian motion framework, 
which is now the cornerstone of modern mathematical finance. 
This is probably unfortunate, because the continuous time, Gaussian random walk 
misses most of the important ``stylized facts'' of financial markets 
--- fat (power-law) tails, intermittency and long memory, etc. 
Many of the results that hold true for a Gaussian process go awry in reality 
--- for example, the well-known perfect hedge of Black-Scholes, 
that would enable one to sell option contracts without any risk, 
is a figment of the very specific assumptions of the model.
It is all too easy to get carried away with the beauty of a mathematical model, 
and forget that it does not bear any relation with reality. 
This is of special concern in the case of financial markets, 
where inadequate models can contribute (and have contributed) to systemic risks \cite{dermanwilmott2009manifesto}. 

This is however not to say that probabilistic methods are useless in this context. 
Quite on the contrary, empirically motivated, faithful models do help 
in controlling risks better and pricing derivative contracts more accurately. 
Many questions that are relevant in practice can be addressed. 
Some of them are directly related to first passage time problems, which is our topic here. 
For example, one might be interested to invest in financial markets with a profit objective, 
that would allow another project to be financed. 
What is the distribution of the time one should wait until this profit is reached? 
Conversely, one might be worried about the default of a company or a bank. 
This is often modeled as the first passage time when a random walk (the value of the asset) 
goes below a certain threshold (the equity) \cite{merton1974pricing}. 
Another example is that of ``barrier options'', which disappear when the price of the underlying 
hits a certain predefined value (the barrier); 
a related issue is the early exercise of so-called American options, 
when the option first reaches a value where it is optimal to exercise and cash the current pay-off 
rather than let the option run to maturity \cite{hull2009options}. 

We review here three examples of the use of first passage ideas in finance:
the design of a goodness-of-fit test whose law is the survival probability of a process before hitting a barrier, 
the optimal time to sell an asset, and the optimal value of the price threshold at which the expected benefit outweighs linear transaction costs.
Extensions and open questions related to these three problems are briefly discussed in the conclusion. 



\clearpage\section{Weighted {K}olmogorov-{S}mirnov tests and first passages~\cite{chicheportiche2012weighted}}
\index[aindx]{Bouchaud, J.-P.} 
\index[aindx]{Chicheportiche, R.} 
Our first example concerns goodness-of-fit (GoF) testing, which is ubiquitous in all fields of science and engineering.
This is the problem of testing whether a null-hypothesis theoretical probability distribution
is compatible with the empirical probability distribution of a sample of observations.
GoF tests are designed to assess quantitatively whether a sample of $N$ observations
can statistically be seen as a collection of $N$ realizations of a given probability law, 
or whether two such samples are drawn from the same hypothetical distribution.

The best known theoretical result is due to Kolmogorov and Smirnov (KS) \cite{kolmogorov1933sulla,smirnov1948table}, 
and has led to the eponymous statistical test for an \emph{univariate} sample of \emph{independent} draws.
The major strength of this test lies in the fact that the asymptotic distribution of its test statistics 
is completely independent of the null-hypothesis cdf.

Several specific extensions have been studied (and/or are still under scrutiny), including: 
different choices of distance measures, 
multivariate samples, 
investigation of different parts of the distribution domain, 
dependence in the successive draws (which is particularly important for financial applications), etc. 

This class of problems has a particular appeal for physicists since the works of Doob \cite{doob1949heuristic} and Khmaladze \cite{khmaladze1982martingale},
who showed how GoF testing is related to stochastic processes. 
Finding the law of a test amounts to computing a survival probability in a diffusion system. 
In a Markovian setting, this is often achieved by treating a Fokker-Planck problem,
which in turn maps into a Schr\"odinger equation for a particle in a certain potential confined by walls.

\subsection{Empirical cumulative distribution and its fluctuations}\label{sec:GoFintro}

Let $\vect{X}$ be a random vector of $N$ independent and identically distributed variables, 
with marginal cumulative distribution function (cdf) $F$. 
One realization of $\vect{X}$ consists of a time series $\{x_1,\ldots,x_n, \ldots, x_N\}$ that exhibits no persistence (see 
Ref.~\cite{chicheportiche2011goodness} when some non trivial dependence is present).
The empirical cumulative distribution function 
\begin{equation}\label{eq:F_N}
    F_N(x)=\frac{1}{N}\sum_{n=1}^N\1{X_n\leq x}
\end{equation}
converges to the true CDF $F$ as the sample size $N$ tends to infinity.
For finite $N$, the expected value and fluctuations of $F_N(x)$ are
\begin{align*}
	\esp{F_N(x)}&=F(x),\\
    \mathrm{Cov}(F_N(x),F_N(x'))&=\frac{1}{N}\left[F(\min(x,x'))-F(x)F(x')\right].
\end{align*}
The rescaled empirical CDF
\begin{equation}\label{eq:Y_N}
    Y_N(u)=\sqrt{N}\,\left[F_N(F^{-1}(u))-u\right]
\end{equation}
measures, for a given $u\in[0,1]$, the difference between the empirically determined cdf 
of the $X$'s and the theoretical one, evaluated at the $u$-th quantile.
It does not shrink to zero as $N\to\infty$, and is therefore the quantity on which any statistics for GoF testing is built.

\subsubsection*{Limit properties}
One now defines the process $Y(u)$ as the limit of $Y_N(u)$ when $N \to \infty$.
According to the Central Limit Theorem, it is Gaussian and its covariance function is given by:
\begin{equation}\label{eq:Itheo}
	{I}(u,v)=\min(u,v)-uv,
\end{equation}
which characterizes the so-called Brownian bridge, i.e.\  a Brownian motion $Y(u)$  such that \mbox{$Y(u\!=\!0)=Y(u\!=\!1)=0$}. 
Interestingly, the function $F$ does not appear in Eq.~(\ref{eq:Itheo}) anymore, 
so the law of any functional of the limit process $Y$ is independent of the law of the underlying finite size sample.
This property is important for the design of \emph{universal} GoF tests.

\subsubsection*{Norms over processes}
In order to measure a limit distance between distributions, a norm $||.||$ over the space of continuous bridges needs to be chosen. 
Typical such norms are the norm-2 (or `Cramer-von Mises' distance)
\[
	||Y||_2=\int_0^1Y(u)^2\d{u},
\]
as the bridge is always integrable, or the norm-sup (also called the Kolmogorov distance)
\[
	||Y||_\infty=\sup_{u\in[0,1]}|Y(u)|,
\]
as the bridge always reaches an extremal value.

Unfortunately, both these norms mechanically overweight the core values $u \approx 1/2$ and disfavor the tails $u \approx 0,1$:
since the variance of $Y(u)$ is zero at both extremes and maximal in the central value, the major contribution to $||Y||$ indeed
comes from the central region and not from the tails. To alleviate this effect, in particular when the GoF test is intended to investigate 
a specific region of the domain, it is preferable to introduce additional weights and study $||Y\sqrt{\psi}||$ rather than $||Y||$ itself.
Anderson and Darling show in Ref.~\cite{anderson1952asymptotic} that the solution to the problem with the Cramer-von Mises norm and arbitrary weights $\psi$
is obtained by spectral decomposition of the covariance kernel. 
They design an eponymous test \cite{darling1957kolmogorov} with the specific choice of $\psi(u)=1/{I(u,u)}$
equal to the inverse variance, which equi-weights all quantiles of the distribution to be tested.
We analyze here the case of the same weights but with the Kolmogorov distance.

So again $Y(u)$ is a Brownian bridge, i.e.\ a centered Gaussian process on $u\in[0,1]$ with covariance function
${I}(u,v)$ given in Eq.~(\ref{eq:Itheo}).
In particular, $Y(0)=Y(1)=0$ with probability equal to 1, no matter how distant $F$ is from the sample cdf around the core values.
In order to put more emphasis on specific regions of the domain, let us weight the Brownian bridge as follows: 
for given $a\in]0,1[$ and $b\in [a,1[$, we define
\begin{equation}\label{eq:weightedY}
	\tilde{y}(u)=y(u)\cdot\left\{\begin{array}{cl}\sqrt{\psi(u)}&,\,a\leq u\leq b\\0&,\,\text{otherwise.}\end{array}\right.
\end{equation}
We will characterize the law of the supremum $K(a,b)\equiv\sup_{u\in[a,b]}\left|\tilde{y}(u)\right|$:
\[
	\mathcal{P}_{\!{\scriptscriptstyle <}}(k|a,b)\equiv\pr{K(a,b)\leq k}
                                                =\Pr{|\tilde{y}(u)|\leq k, \forall u\in[a,b]}.
\]

\subsection[The equi-weighted {B}rownian bridge]{The equi-weighted {B}rownian bridge:\\ {K}olmogorov-{S}mirnov}\label{ssect:unweightedKS}
In the case of a constant weight, corresponding to the classical KS test, the probability 
\mbox{$\mathcal{P}_{\!{\scriptscriptstyle <}}(k;0,1)$} is well defined and has the well known KS form \cite{kolmogorov1933sulla}:
\begin{equation}\label{eq:KSprob}
    \mathcal{P}_{\!{\scriptscriptstyle <}}(k;0,1) = 1 - 2 \sum_{n=1}^{\infty} (-1)^{n-1} \e^{-2 n^2 k^2},
\end{equation}
which, as expected, grows from $0$ to $1$ as $k$ increases. The value $k^*$ such that this probability is $95 \%$ is $k^* \approx 1.358$ 
\cite{smirnov1948table}. This can be interpreted as follows: if, for a data set of size $N$, the maximum value of $|Y_N(u)|$ is larger
than $\approx 1.358$, then the hypothesis that the proposed distribution is a ``good fit'' can be rejected with $95 \%$ confidence.

\subsubsection*{Diffusion in a cage with fixed walls}

The Brownian bridge $Y$ is nothing else than a Brownian motion with imposed terminal condition, and can be written as
$Y(u)=X(u)-u\,X(1)$ where $X$ is a Brownian motion.
The survival probability of $Y$ in a cage with absorbing walls can be found by counting the number of Brownian paths that
go from $Y(0)=0$ to $Y(1)=0$ without ever hitting the barriers.
More precisely, the survival probability of the Brownian bridge in the same stripe can be computed as $f_1(0;k)/f_1(0;\infty)$,
where $f_u(y;k)$ is the transition kernel of the Brownian motion within the allowed region $[-k,k]$. 
It satisfies the simple Fokker-Planck equation
\[
    \left\{
    \begin{array}{rl}
        \partial_uf_u(y;k)&=\displaystyle\frac{1}{2}\partial_y^2f_u(y;k)\\
        f_u(\pm k;k)&=0
    \end{array}
    \right. ,\quad\forall u\in[0,1].
\]
By spectral decomposition of the Laplacian, the solution is found to be
\[
    f_u(y;k)=\frac{1}{k}\sum_{n\in\mathds{Z}}\e^{-E_n u}\,\cos\left(\sqrt{2E_n}\, y\right), \quad\text{where}\quad E_n=\frac{1}{2}\left(\frac{(2n\!-\!1)\pi}{2k}\right)^2
\]
and the free propagator in the limit $k\to\infty$ is the usual
\[
    f_u(y;\infty)=\frac{1}{\sqrt{2\pi u}}\,\e^{-\frac{y^2}{2u}},
\]
so that the survival probability of the constrained Brownian bridge is
\begin{equation}
    \mathcal{P}_{\!{\scriptscriptstyle <}}(k;0,1)=\frac{\sqrt{2\pi}}{k}\sum_{n\in\mathds{Z}}\Exp{-\frac{(2n\!-\!1)^2\pi^2}{8k^2}}.
\end{equation}
Although it looks different from Eq.~\eqref{eq:KSprob}, the two expressions can be shown to be exactly identical.
But the above proof looks to us way easier than the canonical ones \cite{anderson1952asymptotic}.

\subsubsection*{Diffusion in a cage with moving walls}
The problem can be looked at differently. Under the following change of variable and time 
\begin{equation}\label{eq:changeW}
    W(t)=(1+t)\, Y\!\left(\frac{t}{1+t}\right),\quad t=\frac{u}{1-u}\in\left[\frac{a}{1-a},\frac{b}{1-b}\right],
\end{equation}
the problem can be transformed into that of a Brownian diffusion inside a box with walls \emph{moving at constant velocity}. 
Indeed, one can check that
\[
	{\rm Cov}\big(W(t),W(t')\big)=\min(t,t'),
\]
 and that $\mathcal{P}_{\!{\scriptscriptstyle <}}(k|0,1)$ can be now written as
\[
	\mathcal{P}_{\!{\scriptscriptstyle <}}(k|0,)=\Pr{|W(t)|\leq k\,(1+t),\forall t\in[0,\infty[}.
\]
Since the walls expand as $\sim t$ faster than the 
diffusive particle can move ($\sim\sqrt{t}$), the survival probability converges to a positive value, 
which is again given by the usual Kolmogorov distribution \eqref{eq:KSprob} \cite{krapivsky1996life,bray2007survival1,bray2007survival2}.

\begin{figure}
    \includegraphics[scale=.45,trim=0 0 0 25,clip]{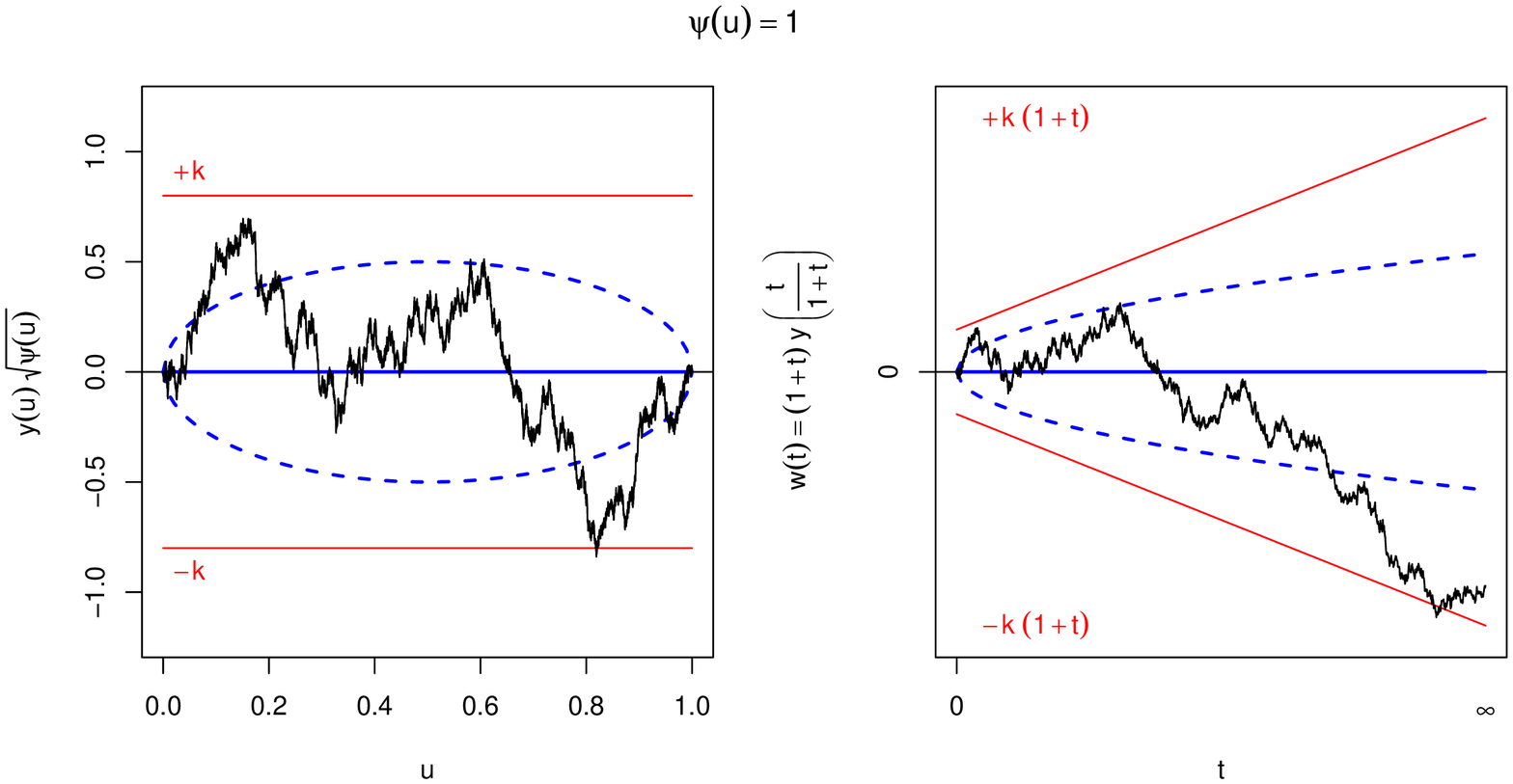}
    \caption{The equi-weighted Brownian bridge, $\psi(u)=1$.
             The time-changed rescaled process lives in a geometry with boundaries receding at constant speed.}\label{fig:BraySmith}
\end{figure}

\subsection[The variance-weighted {B}rownian bridge]{The variance-weighted {B}rownian bridge:\\ Accounting for the tails}\label{ssect:weightedKS}
As mentioned above, the classical KS test is only weakly sensitive to the quality of the fit in the tails of the 
distribution, when it is often these tail events (corresponding to centennial floods, devastating earthquakes, financial crashes, etc.) 
that one is most concerned with (see, e.g., Ref.~\cite{clauset2009power}). 

A simple and elegant GoF test for the tails \emph{only} can be designed starting with digital weights in the form
$\psi(u;a)=\1{u\geq a}$ or $\psi(u;b)=\1{u\leq b}$ for upper and lower tail, respectively.
The corresponding test laws can be read off Eq.~(5.9) in Ref.~\cite{anderson1952asymptotic}.%
\footnote{The quantity $M$ appearing there is the volume under the normal bivariate surface between specific bounds,
and it takes a very convenient form in the unilateral cases $\tfrac12\leq a \leq u \leq 1$ and $0\leq u \leq b\leq\frac12$.
Mind the missing $j$ exponentiating the alternating $(-1)$ factor.}
Investigation of both tails is attained with
$\psi(u;q)=\1{u\leq 1-q}+\1{u\geq q}$ (where $q>\tfrac12$).

Here we rather focus on a GoF test for a univariate sample of size $N\gg 1$, with the Kolmogorov distance but equi-weighted quantiles, 
which is equally sensitive to \emph{all regions} of the distribution.%
\footnote{Other choices of $\psi$ generally result in much harder problems.} 
We unify two earlier attempts at finding asymptotic solutions, one by Anderson and Darling in 1952 \cite{anderson1952asymptotic} 
and a more recent, seemingly unrelated one that deals with ``life and death of a particle in an expanding cage'' 
by Krapivsky and Redner \cite{krapivsky1996life,redner2007guide}. 
We present here the exact asymptotic solution of the corresponding stochastic problem, 
and deduce from it the precise formulation of the GoF test, which is of a fundamentally different nature than the KS test.

So in order to zoom on the tiny differences in the tails of the Brownian bridge, we weight it as explained earlier,
with its variance
\[
	\psi(u)=\frac{1}{u\,(1-u)}.
\]
Solutions for the distributions of such variance-weighted Kolmogorov-Smirnov statistics were studied by No\'e,
leading to the laws of the one-sided \cite{noe1968calculation} and two-sided \cite{noe1972calculation} finite sample tests.
They were later generalized and tabulated numerically by Niederhausen \cite{niederhausen1981tables,wilcox1989percentage}. 
However, although exact and appropriate for small samples, these solutions rely on recursive relations and are not in closed form.
We instead come up with an analytic closed-form solution for large samples that relies on an elegant analogy from statistical physics.

\subsubsection*{Diffusion in a cage with moving walls}

After performing the above change of variable \eqref{eq:changeW} that converts a Brownian bridge into a Brownian motion,
$\mathcal{P}_{\!{\scriptscriptstyle <}}(k|a,b)$ can be written as
\[
	\mathcal{P}_{\!{\scriptscriptstyle <}}(k|a,b)=\Pr{|W(t)|\leq k\sqrt{t},\forall t\in[\tfrac{a}{1-a},\tfrac{b}{1-b}]}.
\]

The problem with initial time $\frac{a}{1-a}=0$ and horizon time $\frac{b}{1-b}=T$ 
has been treated by Krapivsky and Redner in Ref.~\cite{krapivsky1996life}
as the survival probability $S(T;k=\sqrt{\frac{A}{2D}})$ of a Brownian particle 
diffusing with constant $D$ in a cage with walls expanding as $\sqrt{At}$. 
Their result is that for large $T$, 
\[
S(T;k)\equiv\mathcal{P}_{\!{\scriptscriptstyle <}}(k|0,\tfrac{T}{1+T})\propto T^{-\theta(k)}.
\]
They obtain analytical expressions for $\theta(k)$ in both limits \mbox{$k\to 0$} and \mbox{$k\to\infty$}.
The limit solutions of the very same differential problem were found earlier by Turban for the critical behavior of the directed self-avoiding walk in parabolic 
geometries \cite{turban1992anisotropic}.

We take here a slightly different route, suggested (but not finalized) by Anderson and Darling in Ref.~\cite{anderson1952asymptotic}. Our specific contributions are: 
(i) we treat the general case $a>0$ for \emph{any} $k$;
(ii) we explicitly compute the $k$-dependence of both the exponent \emph{and} the prefactor of the power-law decay; and
(iii) we provide the link with the theory of GoF tests and 
compute the pre-asymptotic distribution when $]a,b[\to]0,1[$ of the weighted Kolmogorov-Smirnov test statistics.

\begin{figure}
    \includegraphics[scale=.45,trim=0 0 0 25,clip]{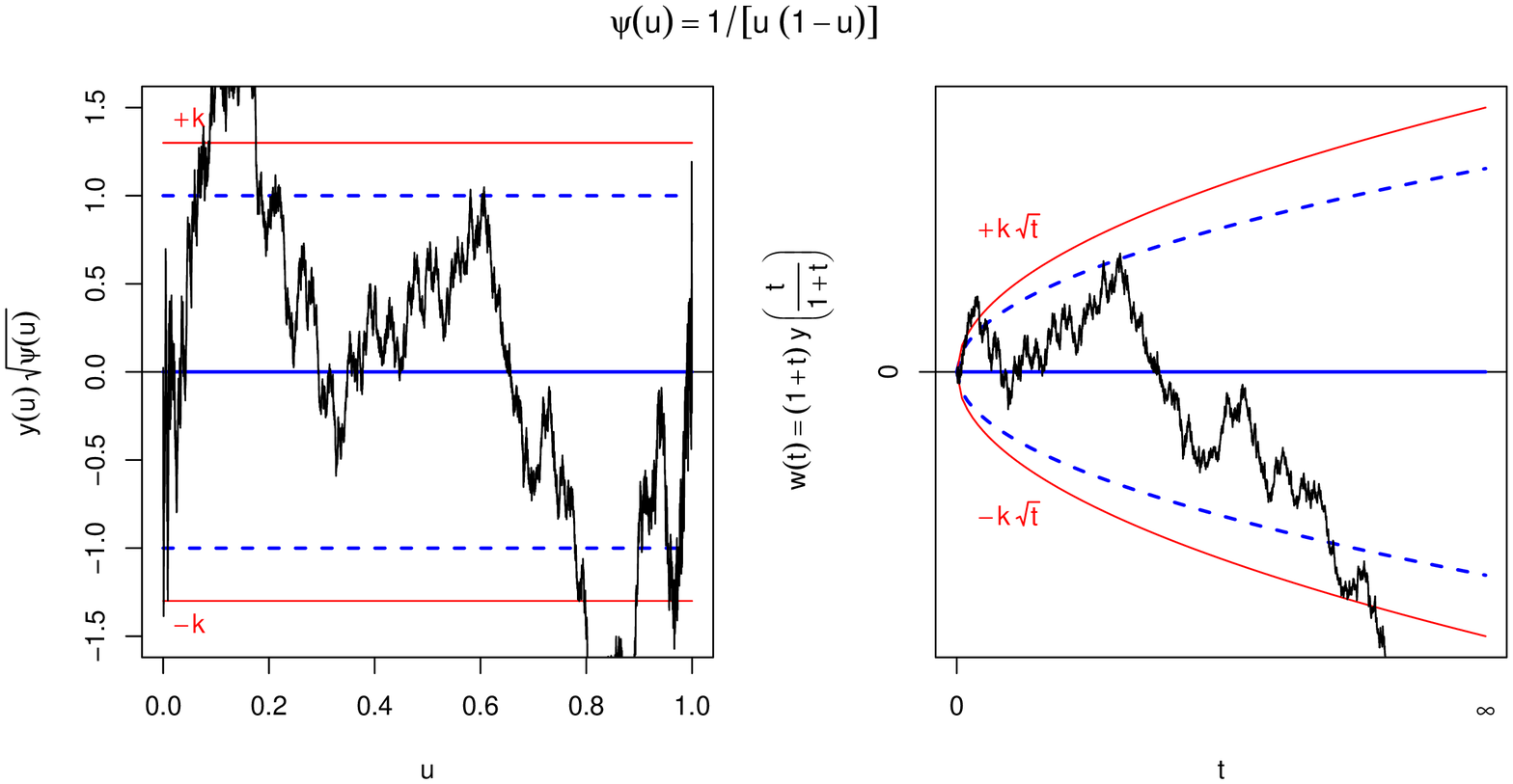}
    \caption{The variance-weighted Brownian bridge, $\psi(u)=1/[u(1-u)]$.
             The time-changed rescaled process lives in a geometry with boundaries receding as $\sim\sqrt{t}$.}\label{fig:KrapRedner}
\end{figure}

\subsubsection*{Mean-reversion in a cage with fixed walls} 

Introducing now the new time change $\tau=\ln\sqrt{\frac{1-a}{a}\,t}$, the variable $Z(\tau)=W(t)/\sqrt{t}$ is a stationary 
Ornstein-Uhlenbeck process on $[0,T]$ 
where 
\begin{equation}\label{eq:Tchange}
    T=\ln\sqrt{\frac{b\,(1-a)}{a\,(1-b)}},
\end{equation}
and
\[
	{\rm Cov}\big(Z(\tau),Z(\tau')\big)=\e^{-|\tau-\tau'|}.
\]
Its dynamics is described by the stochastic differential equation
\begin{equation}
	\d{Z(T)}=-Z(T)\d{T}+\sqrt{2}\,\d{B(T)},
\end{equation}
with $B(T)$ an independent Wiener process.
The initial condition for $T=0$ (corresponding to $b=a$) is $Z(0)=Y(a)/\sqrt{\var{Y(a)}}$,
a random Gaussian variable of zero mean and unit variance. The distribution $\mathcal{P}_{\!{\scriptscriptstyle <}}(k|a,b)$ can now be understood as 
the unconditional survival probability of a mean-reverting particle in a cage with fixed absorbing walls:
\begin{align*}
    \mathcal{P}_{\!{\scriptscriptstyle <}}(k|T)
	      &=\Pr{-k\leq Z(\tau)\leq k, \forall \tau\in[0,T]}\\
	      &=\int_{-k}^{k}f_T(z;k)\,\d{z},
\end{align*}
where 
\[
    f_T(z;k)\,\d{z}=\mathds{P}\big[Z(T)\in[z,z+\d{z}[ \mid \{Z(\tau)\}_{\tau<T}\big]
\]
is the density probability of the particle being at $z$
at time $T$, when walls are in $\pm k$. Its dependence on $k$, although not explicit on the right hand side, is due to the boundary condition 
associated with the absorbing walls (it will be dropped in the following for the sake of readability)%
\footnote{In particular, $\mathcal{P}_{\!{\scriptscriptstyle <}}(k|0)=\operatorname{erf}\left(\tfrac{k}{\sqrt{2}}\right)$.}.

The Fokker-Planck equation governing the evolution of the density $f_T(z)$ reads
\[
	\partial_{\tau}f_{\tau}(z)=\partial_z\left[z\,f_{\tau}(z)\right]+\partial_z^2\left[f_{\tau}(z)\right],\quad 0<\tau\leq T.
\]
Calling $\mathcal{H}_{\text{FP}}$ the second order differential operator $-\left[\mathds{1}+z\partial_z+\partial_z^2\right]$, 
the full problem thus amounts to finding the general solution of
\[
    \bigg\{
	\begin{array}{rcl}
	-\partial_{\tau}f_\tau(z)&=&\mathcal{H}_{\text{FP}}(z)f_\tau(z)\\
	f_{\tau}(\pm k)&=&0, \forall \tau\in[0,T]
	\end{array}
    \bigg. .
\]
We have explicitly introduced a minus sign since we expect that the density decays with time in an absorption problem.
Because of the term $z\partial_z$, $\mathcal{H}_{\text{FP}}$ is not hermitian and thus cannot be diagonalized.
However, as is well known, one can define $f_{\tau}(z)=\e^{-\frac{z^2}{4}}\phi_{\tau}(z)$ and the Fokker-Planck equation becomes
\[
    \bigg\{
	\begin{array}{rcl}
	-\partial_{\tau}\phi_\tau(z)&=&\left[-\partial_z^2+\frac{1}{4}z^2-\frac12\mathds{1}\right]\phi_{\tau}(z)\\
	\phi_{\tau}(\pm k)&=&0, \forall \tau\in[0,T]
	\end{array}
    \bigg. ,
\]
Its Green's function, i.e.\ the (separable) solution \emph{conditionally on the initial position} $(z_{\text{i}},T_{\text{i}})$, 
is the superposition of all modes
\[
	G_{\phi}(z,T\mid z_{\text{i}},T_{\text{i}})=\sum_{\nu}\e^{-\theta_{\nu}(T-T_{\text{i}})}\widehat\varphi_{\nu}(z)\widehat\varphi_{\nu}(z_{\text{i}}),
\]
where $\widehat\varphi_{\nu}$ are the normalized solutions of the stationary Schr\"odinger equation
\[
    \Bigg\{
	\begin{array}{rcl}
	\left[-\partial_z^2+\frac{1}{4}z^2\right]\varphi_{\nu}(z)&=&\left(\theta_{\nu}+\frac12\right)\varphi_{\nu}(z)\\
	\varphi_{\nu}(\pm k)&=&0
	\end{array}
    \Bigg.,
\]
each decaying with its own energy $\theta_\nu$, 
where $\nu$ labels the different solutions with increasing eigenvalues,
and the set of eigenfunctions $\{\widehat\varphi_{\nu}\}$ defines an orthonormal basis of the Hilbert space on which $\mathcal{H}_{\text{S}}(z)=\left[-\partial_z^2+\frac{1}{4}z^2\right]$ acts.
In particular, 
\begin{equation}\label{eq:orthonormality}
	\sum_{\nu}\widehat\varphi_{\nu}(z)\widehat\varphi_{\nu}(z')=\delta(z-z'),
\end{equation}
so that indeed $G_\phi(z,T_{\text{i}}\mid z_{\text{i}},T_{\text{i}})=\delta(z-z_{\text{i}})$, and the general solution writes
\begin{align*}
	f_T(z_T;k)
		       &=\int_{-k}^{k}\e^{\frac{z_{\text{i}}^2-z_T^2}{4}}G_{\phi}(z_T,T\mid z_{\text{i}},T_{\text{i}})\,f_0(z_{\text{i}})\,\d{z_{\text{i}}},
\end{align*}
where $T_{\text{i}}=0$, which corresponds to the case $b=a$ in Eq.~(\ref{eq:weightedY}), and $f_0$ is the distribution of the initial value $z_{\text{i}}$
which is here, as noted above, Gaussian with unit variance.

$\mathcal{H}_{\text{S}}$ figures out an harmonic oscillator of mass $\frac12$ and frequency $\omega=\frac{1}{\sqrt{2}}$ within 
an infinitely deep well of width $2k$: its eigenfunctions are parabolic cylinder functions \cite{mei1983harmonic,gradshteyn1980table}
\begin{align*}
    y_+(\theta;z)&=\phantom{z\,}\e^{-\frac{z^2}{4}}\,\FF{_1}{F}{_1}\!\left(-\tfrac{  \theta}{2},\tfrac{1}{2},\tfrac{z^2}{2}\right)\\
    y_-(\theta;z)&=         z\, \e^{-\frac{z^2}{4}}\,\FF{_1}{F}{_1}\!\left( \tfrac{1-\theta}{2},\tfrac{3}{2},\tfrac{z^2}{2}\right)
\end{align*}
properly normalized. 
The only acceptable solutions for a given problem are the linear combinations of $y_+$ and $y_-$ 
which satisfy orthonormality (\ref{eq:orthonormality})  and the boundary conditions:
for periodic boundary conditions, only the integer values of $\theta$ would be allowed, 
whereas with our Dirichlet boundaries \mbox{$|\widehat\varphi_{\nu}(k)|=-|\widehat\varphi_{\nu}(-k)|=0$}, real non-integer eigenvalues $\theta$ are allowed.%
For instance, the fundamental level $\nu=0$ is expected to be the symmetric solution 
\mbox{$ \widehat\varphi_0(z)\propto y_+(\theta_0;z)$}
with $\theta_0$ the smallest possible 
value compatible with the boundary condition:
\begin{equation}
	\theta_0(k)=\inf_{\theta>0}\big\{\theta:y_+(\theta;k)=0\big\}.
\end{equation}
In what follows, it will be more convenient to make the $k$-dependence explicit, and
a hat will denote the solution with the normalization relevant to our problem, 
namely 
 $\widehat{\varphi}_{0  }(z;k)=y_+(\theta_0(k);z)/||y_+||_k$,
with the norm
\[
    ||y_+||_k^2 \equiv\int_{-k}^ky_+(\theta_0(k);z)^2\,\d{z},
\]
so that
$
    \int_{-k}^k\widehat{\varphi}_{\nu}(z;k)^2\,\d{z}=1.
$

\subsubsection*{Asymptotic survival rate}

Denoting by $\Delta_{\nu}(k)\equiv[\theta_{\nu}(k)-\theta_0(k)]$ the gap between the excited levels and the fundamental, 
the higher energy modes $\widehat\varphi_\nu$ cease to contribute to the Green's function when $\Delta_{\nu}T\gg 1$, 
and their contributions to the above sum die out exponentially as $T$ grows.
Eventually, only the lowest energy mode $\theta_0(k)$ remains, and the solution tends to 
\[
	f_{T}(z;k)=A(k)\,\e^{-\frac{z^2}{4}}\,\widehat\varphi_{0}(z;k)\,\e^{-\theta_{0}(k)T},
\]
when $T\gg (\Delta_1)^{-1}$, with 
\begin{equation}\label{eq:A_tilde}
	A(k)=\int_{-k}^k\e^{\frac{z_{\text{i}}^2}{4}}\widehat\varphi_0(z_{\text{i}};k)f_0(z_{\text{i}})\,\d{z_{\text{i}}}.
\end{equation}

Let us come back to the initial problem of the weighted Brownian bridge reaching its extremal value in $[a,b]$.
If we are interested in the limit case where $a$ is arbitrarily close to $0$ and $b$ close to $1$, then $T \to \infty$ and
the solution is thus given by
\begin{align*}
    \mathcal{P}_{\!{\scriptscriptstyle <}}(k|T)&=A(k)\,\e^{-\theta_{0}(k)T}\int_{-k}^{k}\e^{-\frac{z^2}{4}}\widehat\varphi_{0}(z;k)\,\d{z}\\
            &=\widetilde{A}(k)\, \e^{-\theta_{0}(k)T},  
\end{align*}
with $\widetilde{A}(k) \equiv \sqrt{2\pi}A(k)^2$.

We now compute explicitly the limit behavior of both $\theta_0(k)$ and $\widetilde{A}(k)$.
\paragraph*{$\boxed{k\to \infty}$} 
As $k$ goes to infinity, the absorption rate $\theta_0(k)$ is expected to converge toward $0$: intuitively, an infinitely 
far barrier will not absorb anything. At the same time, $\mathcal{P}_{\!{\scriptscriptstyle <}}(k|T)$ must tend to 1 in that limit. 
So $\widetilde{A}(k)$ necessarily tends to unity. Indeed,
\begin{align}
	       \theta_0(k)&\xrightarrow{k\to\infty} \sqrt{\frac{2}{\pi}}k\,\e^{-\frac{k^2}{2}} \to 0,\\\nonumber
	  \widetilde{A}(k)&\xrightarrow{k\to\infty}\left(\int_{-\infty}^{\infty}\widehat{\varphi}_{0}(z;\infty)^2\,\d{z}\right)^2 = 1.
\end{align}
In principle, we see from Eq.~(\ref{eq:A_tilde}) that corrections to the latter arise both (and jointly) from 
the functional relative difference of the solution 
\mbox{$\epsilon(z;k)=y_+(\theta_0(k);z)/y_+(0;z)-1$}, 
and from the finite integration limits ($\pm k$ instead of $\pm \infty$).
However, it turns out that the correction of the first kind is of second order in $\epsilon$, see \cite{chicheportiche2012weighted}.
The correction to $A(k)$ 
is thus dominated by the finite integration limits $\pm k$, so that
\begin{equation}
    \widetilde{A}(k\to\infty)\approx \operatorname{erf}\left(\frac{k}{\sqrt{2}}\right)^2.
\end{equation}
\paragraph*{$\boxed{k\to 0}$}
For small $k$, the system behaves like a free particle in a sharp and infinitely deep well, since the quadratic potential is almost flat around 0.
The fundamental mode becomes then 
\[
	\widehat\varphi_0(z;k\to 0)=\frac{1}{\sqrt{k}}\cos \left(\frac{\pi z}{2k}\right),
\]
and consequently
\begin{align}
	       \theta_0(k)&\xrightarrow{k\to 0}\frac{\pi^2}{4k^2}-\frac12,\\
	  \widetilde{A}(k)&\xrightarrow{k\to 0}
                        \frac{16}{\pi^2\sqrt{2\pi}}k.
\end{align}

\begin{figure}[!p!]
	\center
	\includegraphics[scale=0.55,trim=0 0 -25 0,clip]{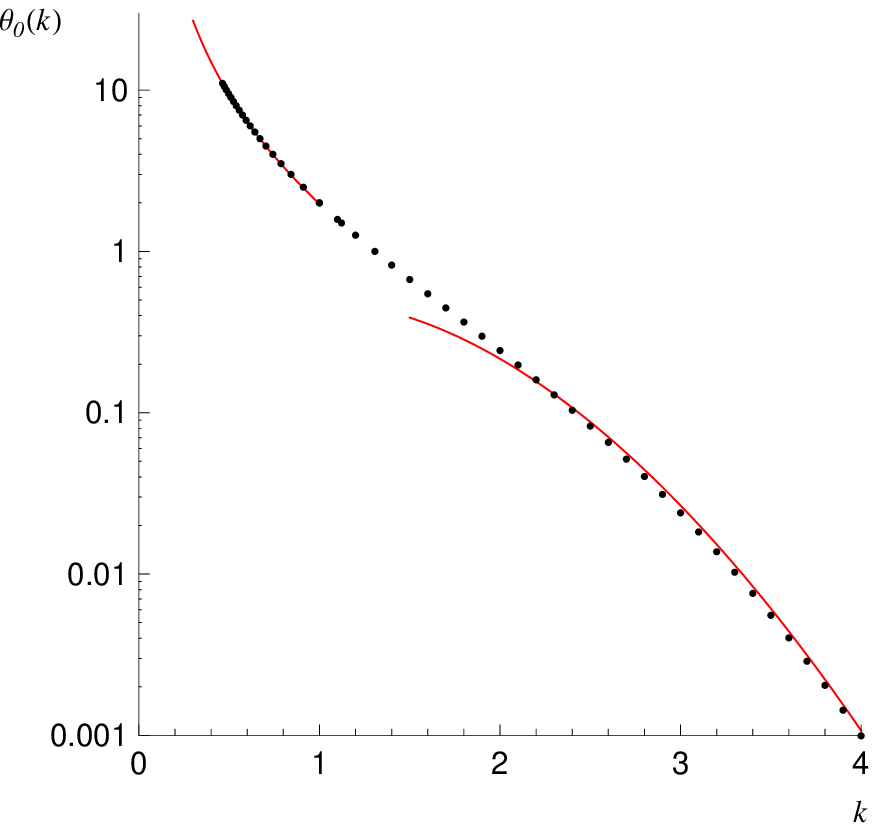}
	\includegraphics[scale=0.55,trim=0 0 -25 0,clip]{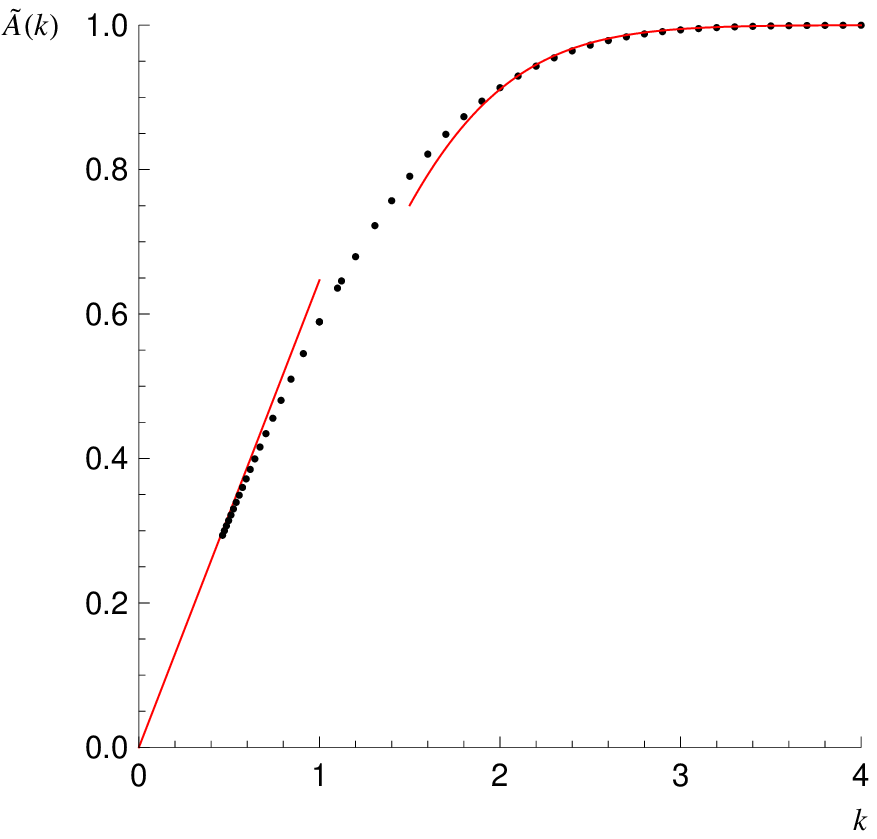}
	\caption{\textbf{Left:  } Dependence of the exponent       $\theta_0$ on $k$; similar to Fig.~2 in Ref.~\cite{krapivsky1996life}, but in lin-log scale; see in particular Eqs.~(9b) and (12) there.
	         \textbf{Right: } Dependence of the prefactor $\widetilde{A}$ on $k$. The red solid lines illustrate the analytical behavior in the
	         limiting cases $k\to 0$ and $k\to\infty$.}
	\label{fig:A_theta}
\end{figure}

We show in Fig.~\ref{fig:A_theta} the functions $\theta_0(k)$ and $\widetilde{A}(k)$ computed numerically from the exact solution,
together with their asymptotic analytic expressions. In intermediate values of $k$ (roughly between 0.5 and 3) these limit
expressions fail to reproduce the exact solution.

\subsubsection*{Higher modes and validity of the asymptotic ($N\gg 1$) solution}
\begin{figure}[!p!]
	\centering
    \minifigure[$1/\Delta_1(k)$ saturates to 1, so that the condition \mbox{$N\gg\exp[1/\Delta_1(k)]$} is virtually always satisfied.]{\includegraphics[scale=0.5,trim=0 0 -25 0,clip]{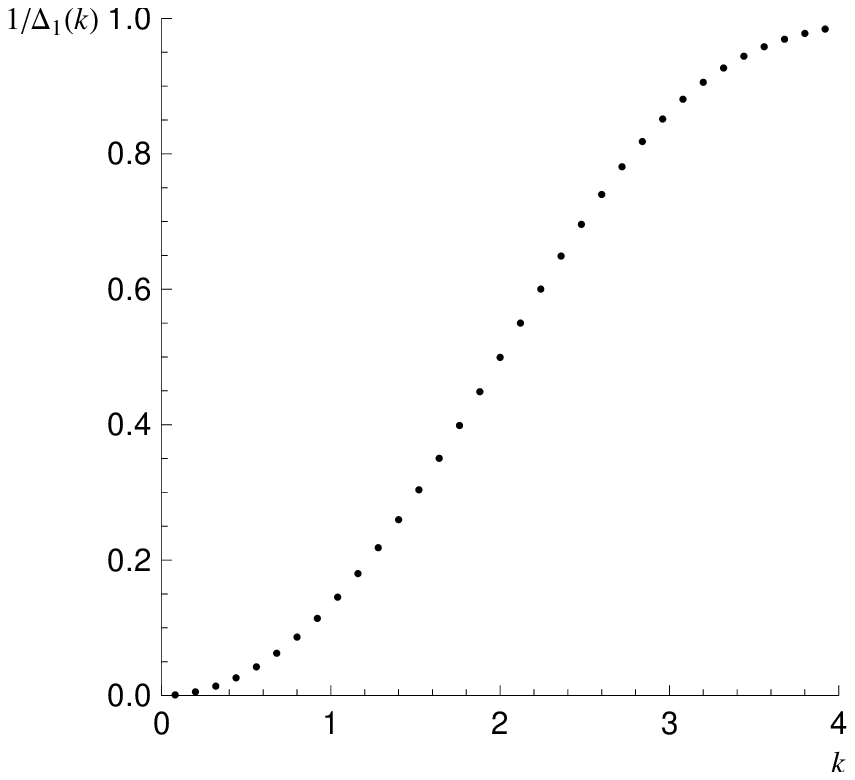}\label{fig:delta1}}
	\minifigure[Dependence of $S(N;k)$ on $k$ for $N=10^3,10^4,10^5,10^6$ (from left to right).
	         The red solid lines illustrate the analytical behavior in the
	         limiting cases $k\to 0$ and $k\to\infty$.]{\includegraphics[scale=0.5,trim=0 10 -25 0,clip]{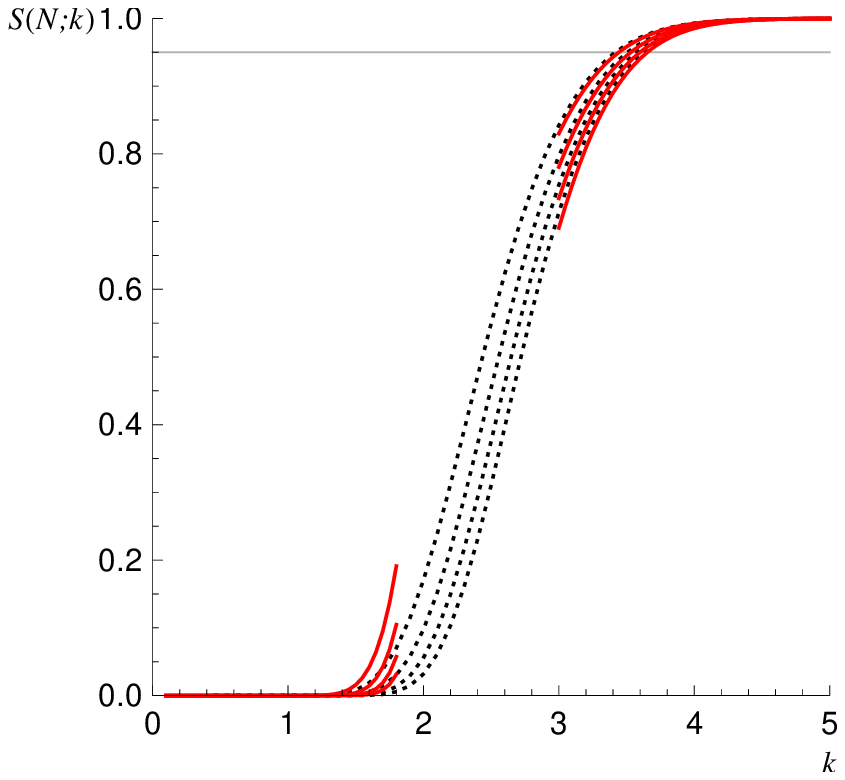}\label{fig:S}}
\end{figure}

Higher modes $\nu>0$ with energy gaps $\Delta_\nu\lesssim 1/T$ must in principle be kept in the pre-asymptotic computation.
This, however, is irrelevant in practice since the gap \mbox{$\theta_1-\theta_0$} is never small.
Indeed, $\widehat\varphi_1(z;k)$ is proportional to the asymmetric solution $y_-(\theta_1(k);z)$
and its energy
\[
    \theta_1(k)=\inf_{\theta>\theta_0(k)}\big\{\theta:y_-(\theta;k)=0\big\}
\]
is found numerically to be very close to \mbox{$1+4\theta_0(k)$}. 
In particular, $\Delta_1>1$ (as we illustrate in Fig.~\ref{fig:delta1}) 
and thus $T \Delta_1 \gg 1$ will always be satisfied in cases of interest. 

\subsubsection{Back to {GoF} testing}

Let us now come back to GoF testing. 
In order to convert the above calculations into a meaningful test, one must specify values of $a$ and $b$. 
The natural choice would be $a=1/N$, corresponding to the min of the sample series since 
$F(\min x_n)\approx F_N(\min x_n)=\frac{1}{N}$. 
Eq.~(\ref{eq:Tchange}) above motivates a slightly different value of $a=1/(N\!+\!1)$ and $b=1-a$, such that 
the relevant value of $T$ is given correspondingly by 
\[
	T=\ln\sqrt{\frac{b\,(1-a)}{a\,(1-b)}} = \ln N.
\]
This leads to our central result for the cdf of the weighted maximal Kolmogorov distance $K(\tfrac{1}{N\!+\!1},\tfrac{N}{N\!+\!1})$ 
under the hypothesis that the tested and the true distributions coincide:
\begin{equation}\label{eq:final_res}
	\boxed{S(N;k)=\mathcal{P}_{\!{\scriptscriptstyle <}}(k|\ln N)=\widetilde{A}(k)N^{-\theta_0(k)}},
\end{equation}
which is valid whenever $N \gg 1$ since, as we discussed above, the energy gap $\Delta_1$ is greater than unity.

The final cumulative distribution function (the test law) is depicted in Fig.~\ref{fig:S} for different values of the sample size $N$.
Contrarily to the standard KS case, this distribution \emph{still depends on $N$}:
as $N$ grows toward infinity, the curve is shifted to the right, and eventually $S(\infty;k)$ is zero for any $k$.
In particular, the threshold value $k^*$ corresponding to a $95 \%$ confidence level (represented as a horizontal grey line) increases with $N$. 
Since for large $N$, $k^* \gg 1$ one can use the asymptotic expansion above, which soon becomes quite accurate, as shown in Fig.~\ref{fig:S}. 
This leads to:
\[
\theta_0(k^*) \approx - \frac{\ln 0.95}{\ln N} \approx \sqrt{\frac{2}{\pi}}k^*\,\e^{-\frac{k^{*2}}{2}},
\]
which gives $k^* \approx 3.439, 3.529, 3.597, 3.651$ for, respectively, $N=10^3,10^4,10^5,10^6$. 
For exponentially large $N$ and to logarithmic accuracy, one has: $k^* \sim \sqrt{2 \ln (\ln N)}$. 
This variation is very slow, but one sees that as a matter of principle, the ``acceptable'' maximal 
value of the weighted distance is much larger (for large $N$) than in the KS case.




\clearpage\section{Optimal time to sell a stock\footnote{Joint work with S.~Majumdar~\cite{majumdar2008optimal}, motivated 
by a paper by Shiryaev, Xu and Zhou~\cite{shiryaev2008thou}.}}
\index[aindx]{Bouchaud, J.-P.} 
\index[aindx]{Majumdar, S.} 

Consider the problem of holding a stock at an initial date $t=0$, 
and hoping to sell it back at a time $t=\tau$ before a deadline $t=T$.
The goal is to find the {\it ex ante} optimal selling time $\tau$, 
i.e.\ take a decision at $t=0$ as of when to sell it in the future for optimal profit. 
If the (log-)price can be modelled as a stationary random walk, this is not a restriction of generality, since at any later time 
the problem is identical with however a reduced horizon. 

\subsection{Minimizing the expected distance to the maximum}
In technical terms, we aim at minimizing the expected (relative) spread 
\[
    S(\tau;T)=\frac{M_T-X_{\tau}}{M_T}
\]
between the instantaneous price $X_\tau$ and the {\it ex post} realized maximum 
over the allowed horizon $M_T=\max\{X_t, t\in[0,T]\}$.

In order for this forward-looking problem to be handled analytically,
we impose that the price process follows a (possibly drifted) geometric Brownian motion $X_t=\e^{x_t}$
with
\[
    \d{x}_t=\mu\,\d{t} +\sigma\,\d{B_t}\quad\text{or}\quad \dot{x}=\mu + \sigma\eta
\]
where $B_t$ in the Stochastic Differential Equation (left) is a Wiener process, 
and $\eta=\frac{\d{B_t}}{\d{t}}$ in the Langevin equation (right) is a standard White Gaussian Noise.
We rewrite the maximum value as $M_T=\e^{m_T}$ with obviously $m_T=\max\{x_t,t\in[0,T]\}$.
The problem is clearly invariant under a shift of both $x_t$ and $m_T$,
so that we can arbitrarily set $x_0=0$.

The optimal time to sell is then defined as the solution of the minimization problem
\[
    \tau^*=\argmin_{\tau\in[0,T]}\esp{\ln S(\tau;T)}=\argmin_{\tau\in[0,T]}\esp{s(\tau;T)}
\]
where $s(\tau;T)=m_T-x_\tau$.
The expectation estimator is clearly inter-temporal, and the probability distribution function $\mathcal{P}_\mu$
of $s(\tau;T)$ can be written in terms of the joint density $f_\mu$ of $(x_\tau,m_T)$ as
\begin{align}\nonumber
\mathcal{P}_\mu(s;\tau,T)&=\int_0^\infty\!\int_{-\infty}^\infty f_\mu(x,m;\tau,T)\,\delta(m-x-s)\,\d{x}\,\d{m}\\\label{eq:pdfspread}
                         &=\int_0^\infty f_\mu(m-s,m;\tau,T)\,\d{m}.
\end{align}
This is equivalent to writing $\mathcal{P}_\mu(s;\tau,T)=\int\d{F_\mu(m-s,m;\tau,T)}$, with 
the (partial) cumulative distribution function
\[
    F_\mu(x,m;\tau,T)=\pr{x_\tau=x,m_T\leq m}
\]
counting the fraction of the paths arriving in $[x,x+\d{x}]$ at time $\tau$ 
among all paths never crossing $m$ from below over the whole horizon $[0,T]$. 
It is expressed in terms of the \emph{causal} propagator $G_\mu$ as
the probability of arriving in $x$ at time $\tau$ without ever hitting $m$,
and then arriving anywhere below $m$ in the remaining time $T-\tau$:
\begin{equation}\label{eq:F}
    F_\mu(x,m;\tau,T)=G_\mu(x,\tau;m)\int_{-\infty}^mG_\mu(x',T-\tau;m-x)\,\d{x'}.
\end{equation}

\begin{figure}
    \includegraphics[scale=.45]{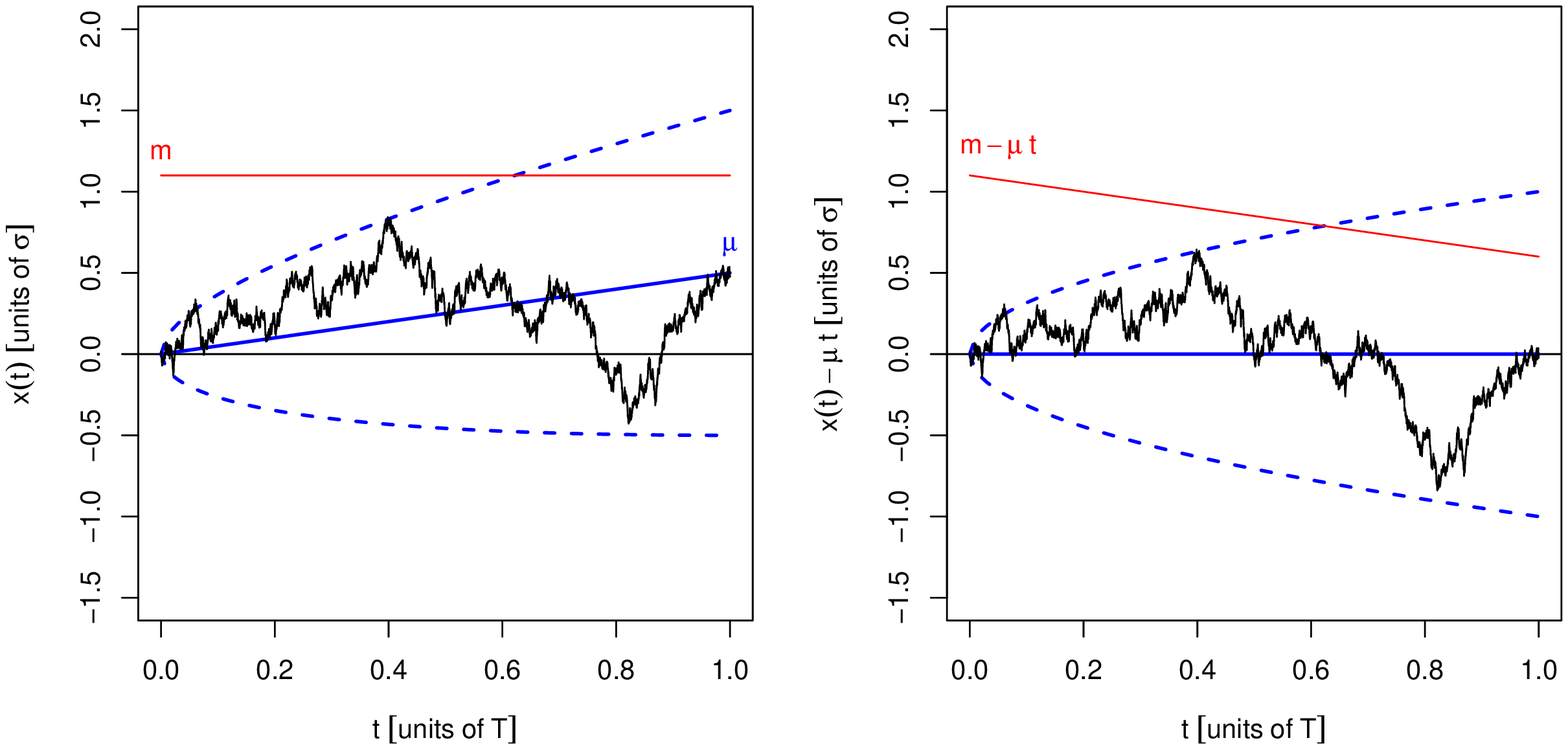}
    \caption{Illustration for a positive drift ($\mu>0$). 
             In the case where $\mu<0$, the upper barrier moves \emph{away} linearly.
             See Sect.~\ref{ssect:unweightedKS} page~\pageref{ssect:unweightedKS} for 
             an application of the bilateral case.}\label{fig:MajBouch}
\end{figure}

The propagator $G_\mu(x,\tau;m)$ describes 
a $\mu$-drifted diffusion close to a fixed absorbing boundary,
or equivalently a pure diffusion close to a boundary moving at constant velocity
(``daredevil at the edge of a receding cliff'', \cite{krapivsky1996life}), see Fig.~\ref{fig:MajBouch}.
It can be written in terms of the propagator $G_0$ of the zero-drift diffusion:
\[
    G_{\mu}(x,t;m)=\Exp{-\frac{\mu^2-2\mu x}{2\sigma^2}}\, G_0(x,t;m),
\]
where $G_0$ can be computed in several ways ---
method of images \cite{feller1968introduction, redner2007guide}, 
path-integral method \cite{majumdar2005brownian, majumdar2008optimal}, 
solution of the Fokker-Planck equation.
The solution writes (up to a normalizing constant) as the difference
$
    G_0(x,t;m)\propto G_0(x,t;\infty)-G_0(x-2m,t;\infty)
$
between the free propagator 
\[
    G_0(x,t;\infty)=\frac{1}{\sqrt{2\pi\sigma^2t}}\,\Exp{-\frac{(x-x_0)^2}{2\sigma^2t}}
\]
with initial positions at $x_0=0$ and $x_0=2m$.
 As expected, $G_0(m,t;m)=0$ at all times, saying that the probability of presence \emph{at} the boundary is nil.

Once $G_\mu$ is known, the joint distribution $f_\mu$ of $x_\tau$ and $m_T$ is obtained by differentiating Eq.~\eqref{eq:F}
with respect to $m$, and the distribution of the spread is found from Eq.~\eqref{eq:pdfspread} to be
\begin{equation}\label{eq:finalP}
\boxed{
    \mathcal{P}_\mu(s;\tau,T)=a_\mu(s;\tau)\,b_{-\mu}(s,T-\tau)+a_{-\mu}(s;T-\tau)\,b_{\mu}(s,\tau)
},
\end{equation}
where
\begin{eqnarray*}
    a_\mu(s;\tau)=&\displaystyle \frac{\mu}{2\sigma^2}&\Exp{-\frac{2s\mu}{\sigma^2}}\Erfc{\frac{s-\mu \tau}{\sqrt{2\sigma^2\tau}}}+
                   \frac{1}{\sqrt{2\pi\sigma^2\tau}}\,\Exp{-\frac{(s+\mu \tau)^2}{2\sigma^2\tau}}\\
    b_\mu(s;\tau)=&\displaystyle -&\Exp{-\frac{2s\mu}{\sigma^2}}\Erfc{\frac{s-\mu\tau}{\sqrt{2\sigma^2\tau}}}+\Erfc{-\frac{s+\mu\tau}{\sqrt{2\sigma^2\tau}}}.
\end{eqnarray*}
Finally, the expected spread is
\begin{equation}\label{eq:exspread}
    \esp{s(\tau,T)}=\int_0^\infty\!s\,\mathcal{P}_\mu(s;\tau,T)\,\d{s}=\iint_0^\infty\!s\,f_\mu(m-s,m;\tau,T)\,\d{m}\,\d{s},
\end{equation}
and the optimal $\tau^*$ is found by minimizing this function with respect to $\tau$:
\begin{equation}\label{eq:t-ast}
    \tau^*= \begin{cases}
                T&, \mu\geq 0\\
                0&, \mu\leq 0
            \end{cases}.
\end{equation}
It is degenerate at $\mu=0$ where both $\tau^*=0$ and $\tau^*=T$ are optimal.
This result states in technical terms a very intuitive truism:
 whenever the log-prices are expected to increase in average ($\mu>0$) one should keep the stock as long as possible,
 and conversely if the log-prices are expected to fall ($\mu<0$) one should sell immediately.
It should not be surprising that $\tau^*$ is not affected by the value of the so-called ``volatility'' parameter $\sigma$,
since the optimization program only focused on ``maximizing the gain'' without controlling for the encountered risk,
and thus the solution applies to a risk-neutral agent only.

\subsection{Maximizing the occurrence time probability of maximum}
Alternatively to minimizing the spread $s(\tau;T)$ between the expected maximum $m_T$ and the log-price $x_\tau$ at the selling time, 
one can try to maximize the probability $p_\mu(\tau;T)\d{\tau}=\d{\pr{x_\tau=m_T}}$ that $m_T$ will occur at time $\tau$, 
whatever value it takes.
But the joint probability that the global maximum over $[0,T]$ has a value $m_T=m$, 
and that this maximum is reached for the first (and in fact only) time at $t=\tau$ is nothing else than $f_\mu(m,m;\tau,T)$.%

In order to avoid the issue of infinite crossings of the continuous Brownian motion \cite{randon2007distribution,majumdar2008time,majumdar2008optimal},
we allow for an infinitesimal spread $s=m-x$ that we eventually take to $0$:
\begin{align*}
    p_\mu(\tau;T)&\propto \int_0^\infty\lim_{s\to0}f_\mu(m-s,m;\tau,T)\,\d{m}=\mathcal{P}_\mu(0^+;\tau,T),
\end{align*}
with $\mathcal{P}_\mu$ given in Eq.~\eqref{eq:finalP}. 
Noticing that $a_\mu$ converges to a finite value $a_\mu(0,\tau)$ when $s\to 0$,
and that the expansion of $b_\mu$ to first order in $s$ around $0$ is $b_\mu(s\to0,\tau)\approx 4s\,a_\mu(0,\tau)$,
we have 
\[
    \mathcal{P}_\mu(s\to0;\tau,T)=8s\, a_\mu(0,\tau)\,a_{-\mu}(0,T-\tau)
\]
and finally, normalizing with $\int_0^T\mathcal{P}_\mu(s;\tau,T)\d{\tau}$, we get
\begin{equation}\label{eq:finalp}
\boxed{
    p_\mu(\tau;T)=2\sigma^2\,a_\mu(0,\tau)\,a_{-\mu}(0,T-\tau)
}.
\end{equation}
Notice that when $\mu=0$ the function $a_0(s,\tau)$ is the centered normal distribution with variance $\sigma^2\tau$.
In particular $a_0(0,\tau)=1/\sqrt{2\pi\sigma^2\tau}$ and one recovers L\'evy's result \cite{levy1939certains}:
\[
    p_0(\tau;T)=\frac{1}{\pi}\frac{1}{\sqrt{\tau\,(T-\tau)}},
\]
with two global maxima at $\tau^\star=0$ and $\tau^\star=T$. For non zero $\mu$'s, the distribution still have inverse
square-root singularities both at $\tau=0$ and $\tau=T$, but with unequal amplitudes. For $\mu < 0$, the amplitude of 
the $\tau = 0$ singularity is larger than that of the $\tau=T$ singularity, and vice-versa when $\mu > 0$. Therefore one 
concludes that:
\begin{equation}
    \tau_m= \argmax_{\tau\in[0,T]}\,p_\mu(\tau;T)
             =\begin{cases}
                 T&, \mu\geq 0\\
                 0&, \mu\leq 0
             \end{cases}
\end{equation}
and is thus equal to $\tau^*$, see Eq.~\eqref{eq:t-ast}.

Whereas the minimization program of the previous section embedded the information of all the possible gaps $s$ 
(as revealed by Eq.~\eqref{eq:exspread}), maximizing the occurrence time distribution only cares for the infinitesimal 
proximity of the maximum $s=m-x\to0$.
Nevertheless, although the objective function is not the same (minimize the spread or maximize the probability),
the solution of the optimal time is not sensitive to the chosen criterion.

\clearpage
\newcommand{\Price}{\textrm{Price}}
\newcommand{\linearise}{\mathfrak{L}}
\newcommand{\piSol}{\pi^*}
\newcommand{\pathIntPred}{\mathcal{G}}
\newcommand{\probaCross}{\mathcal{P}_-}
\newcommand{\diffPos}{\Delta\pi}

\section{Optimal trading with linear costs\footnote{Joint work with J. de~Lataillade, C. Deremble and M. Potters \cite{delataillade2012optimal}.}}
\index[aindx]{Bouchaud, J.-P.} 
\index[aindx]{de~Lataillade, J.} 
\index[aindx]{Deremble, C.} 
\index[aindx]{Potters, M.} 

The problem addressed in this section is to determine
the optimal strategy in the presence of ``linear'' trading costs (i.e. a fixed cost {\it per share}, neglecting any price impact) and a
constraint on the maximum size of the position (both long and short). 
This problem is of very significant interest in practice, at least for small sizes. For large sizes, a quadratic cost can be added to mimic 
price impact; the problem is however not (yet ?) solved in full generality.

We consider an agent who wants to maximize his/her expected gains, 
by trading a single asset, of current price $\Price_t$, over a long period $[0, T]$ 
(we will later consider the limit $T\rightarrow\infty$).
The position (signed number of shares/contracts) of the trader at time $t$ is $\pi_t$. 
We assume that the agent has some signal $p_t$ that predicts the next price change 
$r_t = \Price_{t+1}-\Price_t$, and is faced with the following constraints:
\begin{itemlist}
\item His/her risk control system is simply a cap on the absolute size
  of his/her position~: $|\pi_t|\leq M$, with no other risk control. 
\item He/she has to pay linear costs $\Gamma |\Delta{\pi_t}|$ whenever he/she trades a
  quantity $\Delta{\pi_t}\equiv \pi_{t+1}-\pi_t$
\end{itemlist}\label{linearPred}
We assume that the predictor has a number of ``nice'' (but natural) properties; in particular, 
the predictability $\linearise_{t}(p)=\esp{r_{t}|p_t=p}$ is an odd, continuous and strictly increasing 
function of $p$. We also assume that it is Markovian: $\forall\omega_{t+1}$, $\pr{\omega_{t+1}|p_t,p_{t-1},\dots}=\pr{\omega_{t+1}|p_t}$ 
    where $\omega_{t+1}$ is any event at $t+1$. In what follows, we will use the notation
$P_t(p'|p)\d{p'}= \pr{p_{t+1}\!=\!p' | p_t\!=\!p}\d{p'}$. We also define
  an {integrated predictability} at $t=\infty$, depending on
  $p_t$~:
$$p_\infty(p_t)=\esp{ \Price_{\infty}-\Price_t | p_t } =\sum_{n=0}^\infty \esp{ r_{t+n} | p_t }.$$
This quantity indicates how much one will gain in
the future if one keeps a fixed position $\pi_{t' \ge t}=\pi$~: the
expected gain is then $p_\infty(p_t)\ \pi$.

\subsection{The Optimal Strategy}\label{optimalStrat}

\subsubsection*{A na\"\i ve solution}\label{naive}

At first sight, the solution to this problem seems straightforward: if
the expected future gain (given by the integrated predictability)
exceeds the trading cost per contract $\Gamma$, then one trades in the
direction of the signal (if not already at the maximum position), otherwise one
does not. This solution obviously generates a positive average gain,
but it has no reason to be the optimal solution. Indeed, because the
predictor is auto-correlated in time, it might be worthy (and in
general it will be) to wait for a larger value of the predictor, in
order to grab the opportunities that have the most chances to get
realized, and discard the others.  As we shall see, the mistake in
this na\"\i ve reasoning is not to compare the future gain with the
cost, but rather comes from a wrong definition of the future gain,
which does not include future trading decisions.

\subsubsection*{The Bellman method: general solution}\label{solutionT}

The framework to attack this problem is Bellman's {optimal control theory},
or dynamic programming~\cite{bellman2003dynamic}, 
which consists in solving the problem backwards: 
by assuming one follows the optimal strategy for all future times $t'>t$, 
one can find the optimal solution at time $t$. 
As is usual in dynamic programming, one has a {\it control variable} $\pi_t$, 
which needs to be optimized, and a {\it state variable} $p_t$, 
which parameterizes the solution.
The optimization is done through a {\it value function} $V_t(\pi, p)$, 
which gives the maximal expected gains between time $t$ and $+\infty$, 
considering that the position at $t-1$ is $\pi$ and the predictor's value at $t$ is $p$. 
The optimal solution of the system will be denoted $(\piSol_t)_{t\in[0, T]}$.

At the last time step $t=T$, the expected future return is really $p_\infty(p) \pi_T$ where $p=p_T$, 
since no trading is allowed beyond that time. 
Any trade $\Delta\pi_T$ induces a cost $\Gamma|\Delta\pi_T|$, so:
\begin{unnumlist}
\item If $\phantom{|}p_\infty(p)\phantom{|} \geq + \Gamma$ 
    then $\piSol_T=+M$,
    and  $V_T(\pi,p)=+p_\infty(p) M - \Gamma\,(M\! -\! \pi)$
\item If $\phantom{|}p_\infty(p)\phantom{|} \leq - \Gamma$
    then $\piSol_T=-M$,
    and  $V_T(\pi,p)=-p_\infty(p) M - \Gamma\,(M\! +\! \pi)$
\item If $|p_\infty(p)|<\phantom{-}\Gamma$
    then $\piSol_T=\pi\phantom{M}$,
    and  $V_T(\pi,p)=\phantom{-}p_\infty(p)\pi$.
\end{unnumlist}
Hence, one recovers exactly the na\"\i ve solution in this case, 
but this is only because there is no trading beyond $t=T$.
Now at $t<T$, the quantity to be maximized includes immediate gains, 
costs and future gains. This leads to the following recursion relation:
\begin{gather}\label{basicRecurrence}
  V_t(\pi,p)=\underset{|\pi'|\leq M}{\max}\left\{
    \linearise_{t}(p)\cdot\pi'
  - \Gamma|\pi'-\pi|
  + \int V_{t+1}(\pi',p') P_t(p' |p)  \d{p'}
  \right\},
\end{gather}
and $\pi^*_t$ is the value of $\pi'$ which realises this maximum when $\pi=\piSol_{t-1}$
and $p=p_t$. The general solution is given by the following construction \cite{delataillade2012optimal}:

\begin{itemlist}
\item $\piSol_t=\begin{cases}
                    \piSol_{t-1}&\textrm{if }|p_t|<q_t\\
             \sign(p_t)\cdot M&\textrm{if }|p_t|\geq q_t
       \end{cases}$ \qquad (with $\pi_{-1}=0$)
\item $q_t$ is such that $q_t\geq 0$ and $g(t, q_t)=\Gamma$, where $g(t, p)$ is a continuous, strictly increasing function of $p$
  which satisfies, for $t<T$:
\begin{align}\label{equation_g}
  g(t, p) =\linearise_{t}(p)
         &+\Gamma\left[\int_{q_{t+1}}^\infty-\int_{-\infty}^{-q_{t+1}}\right]P_t(p'|p)\d{p'}\\\nonumber
         &+\phantom{\Gamma\Big[}\int_{-q_{t+1}}^{q_{t+1}}g(t+1,p')P_t(p'|p)\d{p'}
\end{align}
\end{itemlist}

\subsubsection*{The stationary solution}\label{selfCons}

The solution provided by Bellman's method above exhibits in general a dependence in $t$.
Let us now consider the case where $T\rightarrow\infty$, and suppose
that the predictor is stationary, i.e.\ $P_t(p'|p)=P(p'|p)$
is independent of $t$.  
Then we obtain a \textit{telescopic} (self-consistent) equation for the one-variable function $g$, 
and the solution for the threshold $\qstar$:

\begin{align}
  \label{selfConsEq}
    g(p)&=\linearise(p)
         +\Gamma\left[\int_{\qstar}^\infty-\int_{-\infty}^{-\qstar}\right]P(p'|p)\d{p'}
         +\int_{-\qstar}^{\qstar}g(p')P(p'|p)\d{p'}\\
  \label{equation_q}
    g(\qstar)&=\Gamma
\end{align}

The optimal solution $\pi_t^*$ to the system is then similar to the general solution,
but with a constant threshold $\qstar$.
Thus, we obtain a very simple trading system, always saturated at $\pm M$, 
with a threshold to decide at each step whether we should revert the position or not.  
This of course looks a lot like the na\"\i ve solution of page~\pageref{naive}. 
The only difference lies in the value of the threshold $\qstar=g^{-1}(\Gamma)$, 
defined by Eqs.~(\ref{selfConsEq},\ref{equation_q}), 
instead of $q_{\textrm{na\"\i ve}}=p_\infty^{-1}(\Gamma)$ for the na\"\i ve solution. 
Intuitively, these equations take our future trading into account, 
whereas the na\"ive solution does not.

If we look closely at Eq.~\eqref{selfConsEq}, its interpretation
becomes transparent: $2M\,g(p)$ is equal
the expected difference in total future profit between
the situation where $\pi=+M$ and the situation where $\pi=-M$. This
difference is made up of:
\begin{itemlist}
\item $\Delta\pi\, \linearise(p)$ which represents the difference in immediate gain
\item $\Delta\pi\,\Gamma \, \pr{p_{t+1}\!>\! \qstar|p_t\!=\!p}$ which represents the loss if
  the current position is $-M$ and in the next time step the predictor
  goes over the positive threshold $\qstar$ (hence $\pi$ will go to $+M$)
\item $\Delta\pi\,\Gamma \, \pr{p_{t+1}\!<\!-\qstar|p_t\!=\!p}$ which represents the loss if
  the current position is $+M$ and in the next time step the predictor
  goes below the negative threshold $-\qstar$ (hence $\pi$ will go to $-M$)
\item $\int_{-\qstar}^{\qstar}P(p'|p)\ 2Mg(p')\ \d{p'}$ which is
  the expected difference in total future profit if, in the next step,
  the predictor remains between the two thresholds (leaving $\pi$
  unchanged).
\end{itemlist}
Since the change of position between $-M$ and $+M$ costs $2M \Gamma $,
it makes sense to compare $2M g(p)$ with it and only trade when $g(p)$
is greater than $\Gamma$. Hence, $g(p)$ can be seen as the ``gain per
traded lot''.

\label{txt:qLambda}
According to Eq.~(\ref{selfConsEq}), $|g(p)|\geq |\linearise(p)|$ 
and with Eq.~(\ref{equation_q}) this implies in particular that
$\linearise(\qstar)\leq\Gamma$. This property is actually rather intuitive:
indeed, if the immediate expected gain was higher than the trading cost, then
there would be no reason not to trade the maximal possible amount.

From here on, we only consider a linear predictor, $\linearise_t(p)=p$.

\subsubsection*{Reformulation as a path integral}\label{pathIntegral}

Although Eq.~(\ref{selfConsEq}) is easy to interpret, it proves
very difficult to solve in concrete cases. 
It can be rewritten by expanding the function $g$:
\begin{align}\nonumber
  g(p)&= p+
    \int_{-\qstar}^{\qstar}p_1P(p_1|p)\d{p_1}
   +\int_{-\qstar}^{\qstar}\!\int_{-\qstar}^{\qstar}
    p_2P(p_2|p_1)P(p_1|p)\d{p_1}\d{p_2}+\dots\\\nonumber
  &+\Gamma\cdot\left[
    \int_{\qstar}^{+\infty}P(p_1|p)\d{p_1}
   +\int_{\qstar}^{+\infty}\!\int_{-\qstar}^{\qstar}
    P(p_2|p_1)P(p_1|p)\d{p_1}\d{p_2}+\dots \right]\\\nonumber
  &-\Gamma\cdot\left[
    \int_{-\infty}^{-\qstar}P(p_1|p)\d{p_1}
   +\int_{-\infty}^{-\qstar}\!\int_{-\qstar}^{\qstar}
    P(p_2|p_1)P(p_1|p)\d{p_1}\d{p_2}+\dots \right]\\\label{pathIntegralFormula}
  &=\pathIntPred(p)+\Gamma\left[\mathcal{P}_+(p)-\probaCross(p)\right],
\end{align}
and can thus be understood as a path integral: 
$\pathIntPred(p)$ can be interpreted as the average over all possible exit times $n$
of the cumulated predictor $\sum_{i=0}^{n-1}p_i$, where the expectation is taken only
over all paths that stay in $[-\qstar,\qstar]$ until $n$.
Similarly $\mathcal{P}_+(p)$ and $\probaCross(p)$ are the probabilities 
for a path starting at $p_0=p$ to exit (at any possible later time) above $\qstar$ or below $-\qstar$, respectively.

\begin{figure}[ht]
\begin{center}\epsfig{file=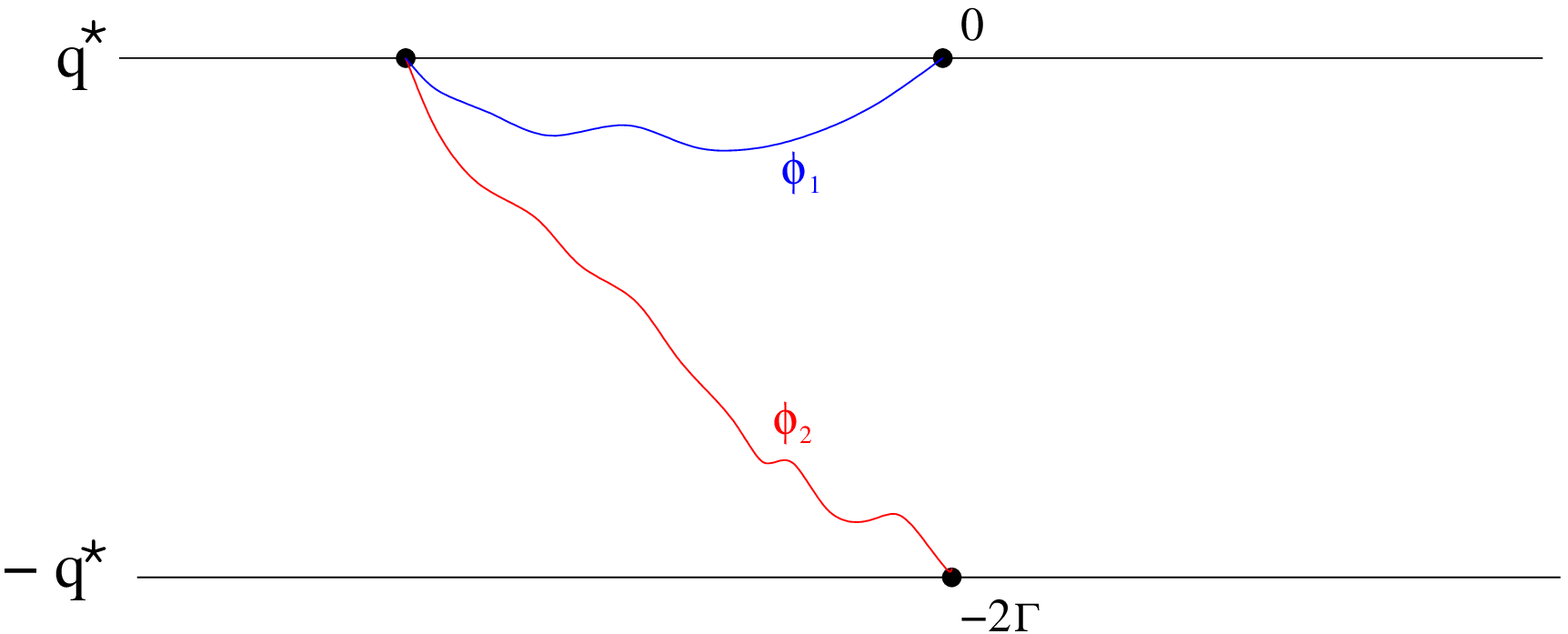, height=4cm}\end{center}
\caption{Path integral representation of Eq.~\eqref{pathIntShort}. 
The value of $\qstar$ is such that the ``penalty'' $2 \Gamma$
over all paths exiting through $-\qstar$ is equal to the average gain
over all paths exiting either through $\qstar$ (e.g.\ $\phi_1$, blue) or through $-\qstar$ (e.g.\ $\phi_2$, red).}\label{pathIntegralFig}
\end{figure}

Using now the fact that $g(\qstar)=\Gamma$, and $\mathcal{P}_+(\qstar)+\probaCross(\qstar)=1$, we get:
\begin{equation}\label{pathIntShort}
    \pathIntPred(\qstar)-2\Gamma\,\probaCross(\qstar)=0.
\end{equation}
In some cases, both sides of this equation will tend to be infinitesimal,
so it is rather the ratio 
$\lim_{p \to \qstar} \pathIntPred(p) / \probaCross(p)$ 
that we will ask to equal $2\Gamma$.
Figure~\ref{pathIntegralFig} illustrates our reformulation of the
problem in terms of first passage times properties.

Note that Eq.~(\ref{pathIntShort}) is completely general provided
the assumptions of page~\pageref{linearPred} are satisfied, it does
not rely on any specific statistics of the predictor.  In the next
section, we will explicitly solve this equation when the predictor is
Gaussian and follows an auto-regressive evolution.

\subsection{Application to an auto-regressive linear predictor}\label{ornstein}

Let us assume that the predictor follows a discrete
auto-regressive dynamics:
\begin{gather}\label{ornsteinUhlenbeck}
p_{t+1}=\rho\cdot p_t+\beta\cdot \xi_t,
\end{gather}
where $(\xi_t)_{t\in\mathbb{R}}$ is a set of independent $\mathcal{N}(0,
1)$ Gaussian random variables.
One classical example of such a
predictor is an exponential moving average of price returns:
$$
p^{\text{EMA}}_t=K\sum_{t'<t}\rho^{t'-t-1}\, r_t.
$$
If we suppose, as is usual, that the returns $r_t$ are 
$\mathcal{N}(0,\sigma_r)$ random variables, then 
$p_t^{\text{EMA}}$ follows the dynamics in Eq.~\eqref{ornsteinUhlenbeck}
 with $\beta=K\sigma_r$. Note however
that the $r_t$ must have some small correlations in order to be
predictable!  Therefore, in this case, the discussion in
terms of an auto-regressive process is only consistent in
the limit  $K \ll 1$.

\medskip
When $\rho=0$, the predictor is a white noise in time: $\esp{ p_tp_{t+1} }=0$.  
Since
we assume a perfect $p \leftrightarrow -p$ symmetry, the self-consistent equation
becomes simply $g(p)=p$, which trivially implies that $\qstar=\Gamma$ in
that case. 
This threshold is what we expect
from such a system: without any auto-correlation, the best strategy is
to trade as soon as the instantaneous predictability is above the
trading cost.  Note that in this case $p_\infty=p_t$, so this
threshold also coincides with the na\"\i ve solution.

\medskip
At the other extreme, when $\epsilon\equiv1-\rho \ll 1$, we have 
$\esp{p_{t+n}|p_t}\approx \e^{-\epsilon n}\,p_t$, 
so $\tau=1/\epsilon$ is the {auto-correlation time} of the predictor $p_t$. 
The standard deviation of the predictor, i.e.\ its {average predictability}, is 
$\sigma_p=\sqrt{\esp{p_t^2}}=\beta/\sqrt{2\epsilon}$.
The integrated predictability is given by
\[
p_\infty(p) =       \sum_{n=0}^\infty \esp{p_{t+n}|p_t} 
            \approx \sum_{n=0}^\infty \e^{-\epsilon n}\,p_t
            \approx p /\epsilon,
\]
what implies that the na\"ive threshold value is given by
$q_{\textrm{na\"ive}}=\Gamma\epsilon$, while the integrated average
predictability is
$\sigma_\infty={\beta}/{\sqrt{2 \epsilon^3}}.$
In what follows, we will study the problem by distinguishing between two cases:
\begin{unnumlist}
\item{$\boxed{\beta\gg\Gamma}$}: the predictor can easily beat its
  transaction costs at every step. This situation (which is not very
  realistic) requires to keep a {\it discrete time} approach of
  the problem.
\item{$\boxed{\beta\ll\Gamma}$}: the predictor needs in general a large
  number of steps to beat the costs. This will lead to a
  {\it continuous} formulation of the problem.
\end{unnumlist}

\subsubsection*{Discrete case:  $\beta\gg\Gamma$}\label{discreteCase}

We already explained in page~\pageref{txt:qLambda} that $\qstar\leq\Gamma$ with a linear predictor.
Consequently, whenever
$\beta\gg\Gamma$, we also have $\beta\gg \qstar$. This means that,
starting at $p=\qstar$, one will typically jump beyond $\qstar$ or $-\qstar$ in
just one step. Thus:
\[
  \pathIntPred(\qstar)=\qstar
  \quad\text{and}\quad
  \probaCross(\qstar)=\int_{x^*}^{+\infty}\frac{\e^{-x^2/2}}{\sqrt{2\pi}} \d{x},
\]
where $x^*={(2-\epsilon)\qstar}/{\beta}$. 
Since $\beta\gg \qstar$, one has $x^*\ll 1$ and thus $\probaCross(\qstar)\approx 1/2$. 
Equation~(\ref{pathIntShort}) finally gives:
$
\qstar=\Gamma.
$
Hence, if the volatility of each predictor change is very large
compared to the trading costs, then one needs to be as selective as
possible.

\subsubsection*{Continuous case: $\beta\ll\Gamma$}\label{continuous}

If the threshold was of the order of the predictor's surprise $\qstar\approx\beta$, 
the predictor would have a significant probability of switching from
above $\qstar$ to below $-\qstar$ in just one step.  
The optimal strategy would then require to resell everything at cost $2\Gamma$, 
whereas the immediate gain would only be of the order of magnitude of $\beta$. 
So when  $\beta\ll\Gamma$, we necessarily have $\qstar\gg\beta$,
and many steps are required for the predictor to get from $\qstar\gg\beta$ to $-\qstar\ll-\beta$.
This is effectively the continuum limit, where the variation of the predictor
at each time step is infinitesimal compared
to $\qstar$. 
We can then approximate the dynamics of the predictor by the Ornstein-Uhlenbeck drift-diffusion process:
\begin{equation}
\label{driftDiffusion}\d{p_t}=-\epsilon\, p_t\,\d{t} + \beta\,\d{X_t},
\end{equation}
where $(X_t)_t$ is a Wiener process.

In such a continuous setting, the quantities $\pathIntPred(\qstar)$ and
$\probaCross(\qstar)$ are actually ill-defined because the diffusion
process starts on an absorbing boundary.  This is a classical problem,
which is handled by starting infinitesimally close to $\qstar$. 
Therefore we consider $\pathIntPred(p)$ and $\probaCross(p)$ for $p =\qstar-\delta< \qstar$.  
It can be shown that these two functions obey two 
{Kolmogorov backward equations}, that read:
\begin{equation}\label{equaDiffL}
    \frac{1}{2}\beta^2 \frac{\partial^2\pathIntPred}{\partial p^2} - \epsilon p \frac{\partial\pathIntPred}{\partial p} = -p 
    \quad\text{and}\quad
    \frac{1}{2}\beta^2 \frac{\partial^2\probaCross }{\partial p^2} - \epsilon p \frac{\partial\probaCross }{\partial p} = 0,
\end{equation}
with boundary conditions: $\pathIntPred(\pm \qstar)=0$ and $\probaCross(\qstar)=0$, $\probaCross(-\qstar)=1$. We therefore
encounter again the problem of a Brownian harmonic oscillator confined between two walls, already discussed in Sect.~\ref{ssect:weightedKS}.
The solution of these equations are
\[
  \pathIntPred(p) =\frac{1}{\epsilon}\left(p-\qstar\frac{I(p\sqrt{a})}{I(\qstar\sqrt{a})}\right) \\
  \quad\text{and}\quad
  \probaCross(p) =\frac{1}{2}\left(1-\frac{I(p\sqrt{a})}{I(\qstar\sqrt{a})}\right),
\]
with
$$
I(x)=\int_0^{x}\e^{v^2}\,\d{v} 
\qquad\textrm{and}\qquad
a=\frac{\epsilon}{\beta^2}.
$$
To first order in $\delta\to0$, Eq.~(\ref{pathIntShort}) becomes
$$
-\frac{\delta}{\epsilon}+\frac{\delta \qstar}{\epsilon}\sqrt{a}\cdot \frac{I'(\qstar\sqrt{a})}{I(\qstar\sqrt{a})}
\approx \Gamma \delta\sqrt{a} \frac{I'(\qstar\sqrt{a})}{I(\qstar\sqrt{a})}.
$$
As expected, $\delta$ disappears from the equation, to give the following
solution for the threshold $\qstar$:
\begin{equation}\label{finalEq}
    \boxed{
    \qstar=\frac{\beta}{\sqrt{\epsilon}} F^{-1}\left(\frac{\Gamma\epsilon^{3/2}}{\beta}\right)
    \quad\textrm{where}\quad 
    F(x)=x-I(x)/I'(x)
    }.
\end{equation}
Note that when $\epsilon\ll 1$, this equation can be expressed
entirely in terms of the integrated predictability:
$$
 p_\infty(\qstar)=\sigma_{\infty}\sqrt{2}\cdot
F^{-1}\left(\frac{\Gamma}{\sigma_{\infty}\sqrt{2}}\right).
$$
This means that we can find the optimal threshold for a predictor by
studying only its total predictive power (if we suppose of course that
it satisfies all the required properties).

One can now study the limits of Eq.~(\ref{finalEq}) for large and
small values of the only remaining adimensional parameter
$\eta={\Gamma\epsilon^{3/2}}/{\beta}$. Interestingly, $\eta \approx 1$ is
the regime of practical interest where predictability beats costs whenever the
predictor's value is of the order of its rms.
The asymptotes of $F(x)$ are as follows:
\begin{itemlist}
\item{If $x \gg 1$}, then $\int_0^x\e^{v^2}\d{v} \ll \e^{x^2}$, so  $F(x) \approx x$.
\item{If $x \ll 1$}, then $F(x) \simeq x-(1-x^2)\int_0^x(1-v^2)\d{v} \approx {2x^3}/{3}$.
\end{itemlist}
Therefore when $\eta \gg 1$, the threshold is simply given by $\qstar =
\Gamma \epsilon$. This result is rather intuitive: if $\beta$ is very
small then the predictability of the predictor is weak, compared to
the trading cost. Hence, it makes sense to try to catch any profitable
opportunity, without taking future trading into account.  That is why
we recover the na\"\i ve solution of page~\pageref{naive}.
If on the other hand $\beta \gg \Gamma\epsilon^{3/2}$ then $\eta \ll 1$,
 and $F^{-1}(\eta) \approx \sqrt[3]{{3}\eta/{2}}$,
which yields
\[
\qstar=\sqrt[3]{\frac{3}{2}\cdot\Gamma\beta^2}.
\]
This says that if $\beta$ is large enough,  the optimal threshold is independent of the mean-reversion parameter $\epsilon$.
The surprise, however, is the rather unexpected dependence of the threshold $\qstar$ as the 1/3 power of the trading costs. 
This result was obtained in the literature before, in the limit considered here of a continuous time random walk, see Refs.~\cite{rogers2004effect,martin2011mean,martin2012optimal}.
Our formulation is however much more general, and would allow one to treat non Gaussian and non stationnary situations as well.

\clearpage\section{Some open problems}

We presented three very different examples of ``first passage time'' problems coming from quantitative finance. 
Let us discuss some extensions and open issues concerning these three problems. 

As far as the Kolmogorov-Smirnov goodness-of-fit test is concerned, we believe that extensions of this test 
to higher-dimensional, multivariate settings, would be quite interesting. 
More precisely, the concept of ``copulas'' (that describe the correlation structure between dependent variables)
has become an important one in theoretical finance in the recent years. 
For pairs of dependent variables, the copula $C(u,v)$ is an increasing function of both its
arguments, from $[0,1] \times [0,1]$ to $[0,1]$. 
The trivial copula, corresponding to independent variables, is such that $C(u,v)=uv$. 
It turns out that it is always possible to transform an arbitrary copula into the independent one 
by an appropriate change of variables, $(u,v) \to (s,t)$ \cite{chicheportiche2013phd,chicheportiche2013ironing}. 
One can then, in the spirit of KS, test the GoF in a \emph{copula independent} manner. 
The problem boils down to estimating the distribution of the maximum of a pinned ``Brownian sheet'' 
that generalizes the Brownian bridge described above. 
This is still an unsolved problem, but there is a hope that an exact solution can be found. 
Extensions to weights that emphasize the ``tails'' of the copula, similar to our one-dimensional problem above, would be quite interesting too. 

The second problem, concerning the optimal selling time, is interesting from a mathematical/pedagogical point of view, 
but the final result turns out to be quite trivial from a financial point of view. 
A more interesting problem would be to add some correlations in the returns, 
accounting for trends or mean-reversion, for example with an exponentially decaying 
correlation function of the lag that would allow to make the problem Markovian.

Finally, the issue of optimal strategies in the presence of transaction costs would deserve much more attention. 
One particularly relevant endeavor would be to solve the problem in the presence of both linear \emph{and} quadratic costs,
i.e.\ when the cost of a change of position $\Delta \pi$ is of the form $\Gamma |\Delta \pi| + \Gamma' |\Delta \pi|^2$.
The case treated in this review corresponds to $\Gamma'=0$, but in practice \emph{price impact} is very important: 
prices tend to go up when one buys, and down when one sells. 

\bibliographystyle{ws-rv-van}
\clearpage
\bibliography{../biblio_all}

\end{document}